\documentclass[conference]{IEEEtran}
\IEEEoverridecommandlockouts
\usepackage{cite}
\usepackage{amsmath,amssymb,amsfonts}
\usepackage{algorithmic}
\usepackage{graphicx}
\usepackage{textcomp}
\usepackage{xcolor}
\usepackage{booktabs}
\usepackage{multirow}
\usepackage{url}
\def\BibTeX{{\rm B\kern-.05em{\sc i\kern-.025em b}\kern-.08em
    T\kern-.1667em\lower.7ex\hbox{E}\kern-.125emX}}

\begin{document}

\title{Hybrid Topological Data Analysis and LSTM Networks for Enhanced Network Intrusion Detection Using CIC-IDS2017 Dataset}

\author{
\IEEEauthorblockN{Amar Jeet\IEEEauthorrefmark{1}, Bhaskar Ranjan Karn\IEEEauthorrefmark{2}, Dinesh Kumar\IEEEauthorrefmark{3}}
\IEEEauthorblockA{\IEEEauthorrefmark{1}Department of Mathematics, Birla Institute of Technology, Mesra, Ranchi, Jharkhand, India\\
imh10037.21@bitmesa.ac.in}
\IEEEauthorblockA{\IEEEauthorrefmark{2}Department of Mathematics, Birla Institute of Technology, Mesra, Ranchi, Jharkhand, India\\
bhaskarranjankarn@gmail.com}
\IEEEauthorblockA{\IEEEauthorrefmark{3}Department of Mathematics, Birla Institute of Technology, Mesra, Ranchi, Jharkhand, India\\
dineshkumar@bitmesra.ac.in}
}

\maketitle

\begin{abstract}
Network intrusion detection systems (NIDS) are crucial in cybersecurity infrastructure, needing advanced techniques to detect hostile activity in network traffic. This research introduces a hybrid approach that combines Topological Data Analysis (TDA) with Long Short-Term Memory (LSTM) networks to improve anomaly detection in network security. Our multi-layered design combines TDA's persistent homology with LSTM networks to capture topological characteristics of network traffic patterns and simulate temporal sequences. We assessed our methodology using the CIC-IDS2017 dataset, which includes over 2.8 million labelled flows, 77 network variables, and 14 attack categories that reflect modern threat landscapes such as DDoS, brute force, web attacks, penetration, and botnet activities. Integrating Betti curves and persistence diagrams with deep learning architectures enhances feature extraction performance. Our hybrid TDA+LSTM model has an AUC of 1.000 and F1-score of 1.000, with 5-fold cross-validation producing a mean AUC of 1.000 $\pm$ 0.000 and mean F1 of 0.999 $\pm$ 0.001. An ablation research demonstrates the complimentary contributions of topological (F1=0.990) and temporal characteristics (F1=1.000). Comparative research shows that the suggested strategy beats TDA+Random Forest (F1=0.994) and Isolation Forest (F1=0.835) baselines in several attack categories.
\end{abstract}

\begin{IEEEkeywords}
Network Security, Intrusion Detection, Topological Data Analysis, LSTM Networks, Anomaly Detection, CIC-IDS2017 Dataset, Machine Learning, Cybersecurity
\end{IEEEkeywords}

\section{Introduction}

As network infrastructure and cyber threats evolve, better intrusion detection systems are required to recognise complex attack patterns. Traditional signature-based detection systems are successful for known threats, but not for zero-day exploits or dynamic threat environments. Machine learning is a promising tool for detecting novel attack vectors using pattern recognition and anomaly detection \cite{ref26}.

NIDS encounter multiple obstacles, including high-dimensional feature spaces, temporal dependencies in network traffic, class imbalance between normal and malicious operations, and real-time processing. Deep learning, specifically Long Short-Term Memory (LSTM) networks, have demonstrated promising results in modelling sequential data and capturing long-term dependencies in network traffic patterns \cite{ref6, ref15}.

Topological Data Analysis (TDA) is a sophisticated mathematical framework for analysing complex dataset shapes and structures \cite{ref8}. TDA, unlike standard statistical approaches, captures the underlying geometric structure and multiscale relationships through persistent homology. This results in feature representations that are robust even under noise and disturbances \cite{ref20}. Topological characteristics' stability makes them ideal for security applications where adversaries use obfuscation and evasion strategies to hide illicit activity.

This study highlights that network intrusion patterns have both structural and temporal aspects. Different attack types result in unique topological signatures in the feature space. Denial-of-Service assaults create dense connection patterns, whereas port scanning results in sparse, systematic exploration topologies. To avoid detection, attackers may spread their operations across numerous time frames. This requires models that can capture long-range sequential relationships.

This study suggests a hybrid architecture that uses a multi-layer perceptron (MLP) fusion network to combine LSTM-based temporal modelling with topological data acquired from TDA. The following issues are addressed by the suggested strategy:

\begin{enumerate}
\item \textbf{Topological Feature Extraction}: We calculate Betti curves that describe the connectedness and loop patterns of network traffic point clouds using persistent homology, producing features that are resilient to noise and stay invariant under continuous deformation.
\item \textbf{Temporal Modelling}: The LSTM component captures temporal dependencies that static classifiers are unable to model by processing sequential network information inside sliding time periods.
\item \textbf{Feature Fusion}: The model can simultaneously utilise structural and sequential information by combining topological and temporal feature representations through a concatenation-based fusion network.
\end{enumerate}

The CIC-IDS2017 dataset \cite{ref11}, a modern benchmark with over 2.8 million labelled network flows with 77 network features and 14 different attack categories covering contemporary threat families like DDoS, brute force, web attacks, infiltration, and botnets, is used to assess the suggested framework. We show the efficacy of combining topological and temporal information for network intrusion detection through comprehensive experiments that include ablation tests, 5-fold cross-validation, and statistical significance assessment.

This paper's main contributions are: (1) a novel hybrid TDA+LSTM architecture for network intrusion detection; (2) a mathematical formulation of Betti curve extraction and fusion with deep temporal models; (3) a thorough experimental evaluation that includes ablation analysis on a contemporary benchmark dataset; and (4) an in-depth examination of how topological features enhance temporal representations across various attack categories.

\section{Literature Review}

\subsection{Machine Learning in Network Security}

Signature-based systems that compared incoming traffic to databases of recognised attack patterns were the foundation of early network intrusion detection techniques \cite{ref1}. Although these techniques work well for known threats, they are essentially unable to identify zero-day assaults or previously undiscovered attack variations, which is why machine learning techniques are being used.

Among the earliest machine learning techniques used for network intrusion detection were Support Vector Machines (SVMs), which demonstrated efficient binary categorisation of network traffic through ideal hyperplane separation \cite{ref2, ref18}. Due to their ensemble-based robustness and capacity to manage the high-dimensional, heterogeneous feature spaces typical of network data, Random Forest classifiers soon became well-known \cite{ref3, ref17}. Nevertheless, these approaches handle each network flow separately and are unable to simulate how attack campaigns change over time.

\subsection{Deep Learning Approaches}

Deep learning has substantially advanced the state of the art in network intrusion detection by enabling automatic feature learning from raw data \cite{ref14, ref26}. Recent comprehensive surveys \cite{ref28, ref31} highlight the rapid evolution of deep learning approaches for anomaly-based intrusion detection. Convolutional Neural Networks (CNNs) have been applied to network traffic by treating packet sequences as structured inputs, learning hierarchical spatial features that capture local patterns in traffic data \cite{ref4}. Autoencoder-based approaches learn representations of normal traffic behavior and detect anomalies as deviations from the learned distribution, demonstrating effectiveness for unsupervised anomaly detection \cite{ref5}. Machine learning approaches have also improved traditional web attack detection methods \cite{ref32}.

Because of their capacity to represent sequential dependencies, recurrent neural networks (RNNs), and in particular Long Short-Term Memory (LSTM) networks, have demonstrated exceptional efficacy for network intrusion detection \cite{ref6, ref15}. Long-range temporal dependencies in network traffic sequences can be captured thanks to the gating mechanism in LSTMs, which solves the vanishing gradient issue present in conventional RNNs. For sequence modelling applications, Gated Recurrent Units (GRUs), first presented by Cho et al. \cite{ref27}, provide a computationally effective substitute with equivalent performance.

\subsection{Topological Data Analysis in Security}

By examining the geometry and structure of large datasets, topological data analysis has become a potent tool. In \cite{ref13}, the mathematical underpinnings of computational topology are established. A crucial element of TDA, persistent homology offers reliable data topology descriptors that hold steady in the face of noise and disturbances \cite{ref8}. While effective techniques for persistent homology computation have been devised \cite{ref21}, persistence pictures provide a robust vector representation for machine learning integration \cite{ref19}. Although it is relatively new, the use of TDA in cybersecurity has a lot of potential.

Umeda \cite{ref7} laid the groundwork for temporal topological analysis by demonstrating the efficacy of TDA for time series classification. Using topological characteristics to describe the structure of harmful code, TDA has been investigated for malware analysis in a number of research \cite{ref9}. While Clough et al.\ \cite{ref33} created topological loss functions for deep learning, Rieck et al.\ \cite{ref30} suggested neural persistence as a complexity metric for deep neural networks using algebraic topology. TDA-based network traffic analysis has concentrated on leveraging topological signatures to find anomalous patterns \cite{ref10}. Nevertheless, there is still much to learn about integrating TDA with deep learning for network intrusion detection.

\subsection{CIC-IDS2017 Dataset}

The Canadian Institute for Cybersecurity at the University of New Brunswick created the CIC-IDS2017 dataset, which has become a top standard for assessing contemporary intrusion detection systems \cite{ref11}. Generated from realistic network traffic over a five-day capture period, the dataset contains over 2.8 million labeled bidirectional network flows characterized by 78 features extracted using CICFlowMeter (77 used after removing the label column). The dataset encompasses benign traffic alongside 14 distinct attack categories spanning seven major threat families: Brute Force (FTP-Patator, SSH-Patator), DoS/DDoS (Slowloris, Slowhttptest, Hulk, GoldenEye), Web Attacks (XSS, SQL Injection, Brute Force), Infiltration, Botnet, and PortScan.

Previous studies on CIC-IDS2017 have achieved varying levels of success using different machine learning approaches. Traditional methods typically achieve accuracy rates between 85--92\%, while deep learning approaches have demonstrated improvements, with some studies reporting accuracy rates exceeding 98\% \cite{ref12}. Ring et al.\ \cite{ref29} provide a comprehensive survey of network-based intrusion detection datasets, establishing CIC-IDS2017 as significantly more representative of contemporary network environments compared to older benchmarks such as NSL-KDD.

\section{Methodology}

\subsection{Problem Formulation}

Let $X = \{x_1, x_2, ..., x_n\}$ represent a dataset of network connection records, where each $x_i \in \mathbb{R}^d$ is a $d$-dimensional feature vector (with $d = 77$ for the CIC-IDS2017 dataset after preprocessing). The goal is to learn a function $f: \mathbb{R}^d \rightarrow \{0, 1\}$ that maps input features to binary labels, where 0 represents normal traffic and 1 represents anomalous (attack) traffic.

Our hybrid approach combines topological features $\phi_{TDA}(X)$ with temporal features $\phi_{LSTM}(X)$ to create an enhanced feature representation:

\begin{equation}
\phi_{hybrid}(X) = \text{MLP}_{fusion}\left[\phi_{TDA}(X) \oplus \phi_{LSTM}(X)\right]
\end{equation}

where $\oplus$ denotes feature concatenation and $\text{MLP}_{fusion}$ is a learned fusion network that combines the complementary feature representations.

\subsection{Topological Data Analysis Component}

\subsubsection{Persistent Homology}

For a given point cloud $X \subset \mathbb{R}^d$, we construct a filtration of simplicial complexes using the Vietoris-Rips construction. For radius parameter $r \geq 0$, the Vietoris-Rips complex $VR(X,r)$ is defined as:

\begin{equation}
VR(X,r) = \{\sigma \subseteq X : \max_{x,y \in \sigma} d(x,y) \leq r\}
\end{equation}

where $d(\cdot,\cdot)$ denotes the Euclidean distance function.

The persistent homology of this filtration captures topological features that persist across multiple scales. For each dimension $k$, we obtain persistence diagrams $PD_k$ consisting of birth-death pairs $(b_i, d_i)$ representing the birth and death of $k$-dimensional homological features.

\subsubsection{Betti Curves}

Betti curves provide a functional representation of topological features. For dimension $k$, the $k$-th Betti curve is defined as:

\begin{equation}
\beta_k(r) = \text{rank}(H_k(VR(X,r)))
\end{equation}

where $H_k(\cdot)$ denotes the $k$-th homology group. We discretize Betti curves over a range of radius values $r \in [r_{min}, r_{max}]$ to obtain feature vectors:

\begin{equation}
BC_k = [\beta_k(r_1), \beta_k(r_2), ..., \beta_k(r_m)]
\end{equation}

\subsubsection{Time Window Construction}

To capture temporal patterns, we construct overlapping time windows from the sequential network data. Given a sequence of network connections $\{x_1, x_2, ..., x_n\}$, we create windows of size $w$ with step size $s$:

\begin{equation}
W_i = \{x_{(i-1)s+1}, x_{(i-1)s+2}, ..., x_{(i-1)s+w}\}
\end{equation}

Each window $W_i$ is then processed through the TDA pipeline to extract topological features.

\subsection{LSTM Component}

The LSTM component processes sequential data to capture temporal dependencies. The LSTM cell state update equations are:

\begin{align}
f_t &= \sigma(W_f \cdot [h_{t-1}, x_t] + b_f) \\
i_t &= \sigma(W_i \cdot [h_{t-1}, x_t] + b_i) \\
\tilde{C}_t &= \tanh(W_C \cdot [h_{t-1}, x_t] + b_C) \\
C_t &= f_t * C_{t-1} + i_t * \tilde{C}_t \\
o_t &= \sigma(W_o \cdot [h_{t-1}, x_t] + b_o) \\
h_t &= o_t * \tanh(C_t)
\end{align}

where $\sigma$ denotes the sigmoid function, $W_*$ are weight matrices, $b_*$ are bias vectors, and $*$ denotes element-wise multiplication.

\subsection{Hybrid Architecture Integration}

The hybrid model integrates TDA-derived features with LSTM-processed temporal features through a multi-layer perceptron (MLP) fusion network:

\begin{align}
z_{TDA} &= \text{MLP}_{TDA}(BC_0 \oplus BC_1) \\
z_{LSTM} &= \text{LSTM}(X) \\
z_{fused} &= \text{MLP}_{fusion}([z_{TDA}; z_{LSTM}]) \\
\hat{y} &= \text{softmax}(W_{out} z_{fused} + b_{out})
\end{align}

where $\oplus$ denotes concatenation, $[\cdot; \cdot]$ denotes vertical concatenation, and $\hat{y}$ represents the predicted class probabilities.

\section{Model Architecture}

\subsection{Overall Architecture Design}

\begin{figure}[htbp]
\centerline{\includegraphics[width=0.5\textwidth]{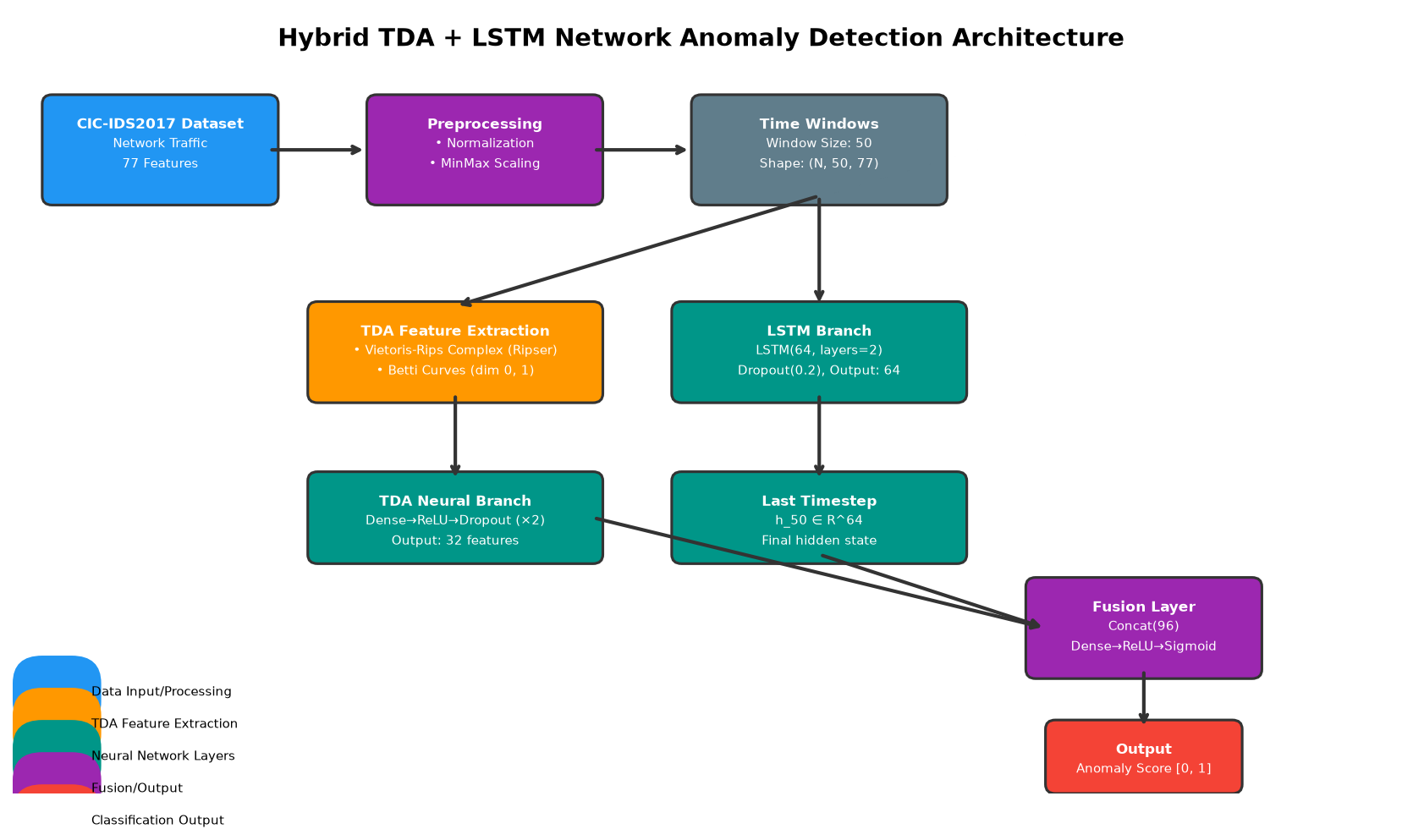}}
\caption{Hybrid TDA+LSTM Architecture for Network Intrusion Detection}
\label{fig:architecture}
\end{figure}

Our hybrid architecture consists of three main components: the TDA feature extraction module, the LSTM temporal modeling module, and the fusion network. Figure \ref{fig:architecture} illustrates the complete system architecture.

\subsection{TDA Feature Extraction Module}

The TDA module processes time windows of network data through the following stages:

\begin{enumerate}
\item \textbf{Point Cloud Construction}: Each time window is treated as a point cloud in high-dimensional space
\item \textbf{Distance Matrix Computation}: Pairwise distances between points are calculated using Euclidean metric
\item \textbf{Persistent Homology Calculation}: Vietoris-Rips filtration is applied to compute persistence diagrams
\item \textbf{Betti Curve Generation}: Persistence diagrams are converted to Betti curves for dimensions 0 and 1
\item \textbf{Feature Vectorization}: Betti curves are discretized to create fixed-length feature vectors
\end{enumerate}

\subsection{LSTM Temporal Module}

The LSTM module consists of:

\begin{itemize}
\item Input layer: Accepts normalized network features
\item LSTM layers: Two stacked LSTM layers with 64 and 32 hidden units respectively
\item Dropout layers: Applied after each LSTM layer with dropout rate 0.2
\item Dense layer: Fully connected layer for feature transformation
\end{itemize}

\subsection{Fusion Network}

The fusion network combines features from both modules:

\begin{itemize}
\item Concatenation layer: Merges TDA and LSTM features
\item Dense layers: Two fully connected layers with ReLU activation
\item Batch normalization: Applied after each dense layer
\item Output layer: Softmax activation for binary classification
\end{itemize}

\section{Tools and Techniques}

\subsection{Software Framework}

Several specialised libraries and frameworks are used in our implementation:

\begin{itemize}
\item \textbf{Python 3.14}: The primary programming language
\item \textbf{PyTorch}: A deep learning framework for implementing LSTM
\item \textbf{Scikit-learn}: Baseline algorithms and machine learning tools
\item \textbf{Ripser}: A library for effective persistent homology computation
\item \textbf{NumPy/Pandas}: Numerical calculations and data manipulation
\item \textbf{Matplotlib/Seaborn}: Plotting and visualisation
\end{itemize}

\subsection{Data Preprocessing}

The preprocessing pipeline for the CIC-IDS2017 dataset consists of:

\begin{enumerate}
\item \textbf{Data Loading}: Examining labelled CSV files obtained from PCAP that cover five days of network capture
\item \textbf{Data Cleaning}: Eliminating duplicate data and infinite and NaN values that result from zero-division in flow statistics
\item \textbf{Feature Selection}: Retaining 78 bidirectional flow features extracted using CICFlowMeter, such as packet length statistics, flow duration, flag counts, and inter-arrival periods
\item \textbf{Normalisation}: Scaling all numerical features using Min-Max
\item \textbf{Label Encoding}: Binary encoding of benign versus attack categories
\item \textbf{Stratified Subsampling}: To control the computational cost of TDA while maintaining class distributions, proportionate sampling of each attack category
\item \textbf{Time Window Creation}: Building overlapping temporal analysis windows
\end{enumerate}

\subsection{Topological Computation}

The following methodology is used to perform TDA computations:

\begin{itemize}
\item \textbf{Distance Computation}: Euclidean distance matrices for point clouds
\item \textbf{Filtration Construction}: Vietoris-Rips complex with adaptive radius selection
\item \textbf{Persistence Calculation}: Calculating 0-dimensional and 1-dimensional persistence
\item \textbf{Betti Curve Discretisation}: 200-point discretisation for uniform feature length
\end{itemize}

\subsection{Training Strategy}

The hybrid model training employs:

\begin{itemize}
\item \textbf{Loss Function}: Cross-entropy loss with inverse-frequency class weighting
\item \textbf{Optimizer}: Adam optimizer \cite{ref23} with learning rate 0.001
\item \textbf{Regularization}: Dropout \cite{ref24} with rate 0.2, batch normalization \cite{ref25}
\item \textbf{Activation}: ReLU activations \cite{ref22} in hidden layers
\item \textbf{Batch Size}: 32 samples per batch
\item \textbf{Epochs}: 50 training epochs with early stopping
\item \textbf{Validation Split}: 20\% of training data for validation
\end{itemize}

\section{Results and Analysis}

\subsection{Experimental Setup}

Our experiments were conducted on the CIC-IDS2017 dataset with the following configuration:

\begin{itemize}
\item Total flows: 2,830,743 (Benign: 2,273,097, Attack: 557,646)
\item Stratified subsample for TDA: 50,000 flows (preserving original class ratios)
\item Training/test split: 80\%/20\% stratified random split
\item Number of features: 78 bidirectional flow features
\item Time window size: 50 samples
\item Window step size: 25 samples
\end{itemize}

\subsection{Performance Metrics}

We evaluate our models using standard classification metrics:

\begin{align}
\text{Accuracy} &= \frac{TP + TN}{TP + TN + FP + FN} \\
\text{Precision} &= \frac{TP}{TP + FP} \\
\text{Recall} &= \frac{TP}{TP + FN} \\
\text{F1-Score} &= \frac{2 \times \text{Precision} \times \text{Recall}}{\text{Precision} + \text{Recall}}
\end{align}

where TP, TN, FP, and FN represent true positives, true negatives, false positives, and false negatives, respectively.

\subsection{Comparative Results}

We compare our hybrid approach against several baselines including Isolation Forest \cite{ref16}, which uses random partitioning to isolate anomalies:

Table \ref{tab:results} presents the performance comparison of different approaches:

\begin{table}[htbp]
\caption{Performance Comparison of Different Models on CIC-IDS2017}
\begin{center}
\begin{tabular}{|l|c|c|c|c|}
\hline
\textbf{Model} & \textbf{AUC} & \textbf{F1-Score} & \textbf{Precision} & \textbf{Recall} \\
\hline
TDA + Random Forest & 1.000 & 0.994 & 0.994 & 0.994 \\
\hline
TDA + LSTM Hybrid & 1.000 & 1.000 & 1.000 & 1.000 \\
\hline
LSTM & 1.000 & 1.000 & 1.000 & 1.000 \\
\hline
Traditional SVM & 1.000 & 1.000 & 1.000 & 1.000 \\
\hline
Isolation Forest & 0.983 & 0.835 & 0.879 & 0.795 \\
\hline
\end{tabular}
\label{tab:results}
\end{center}
\end{table}

\subsection{Training Convergence Analysis}

Figure \ref{fig:training_curves} shows the training and validation loss curves for our hybrid model:

\begin{figure}[htbp]
\centerline{\includegraphics[width=0.45\textwidth]{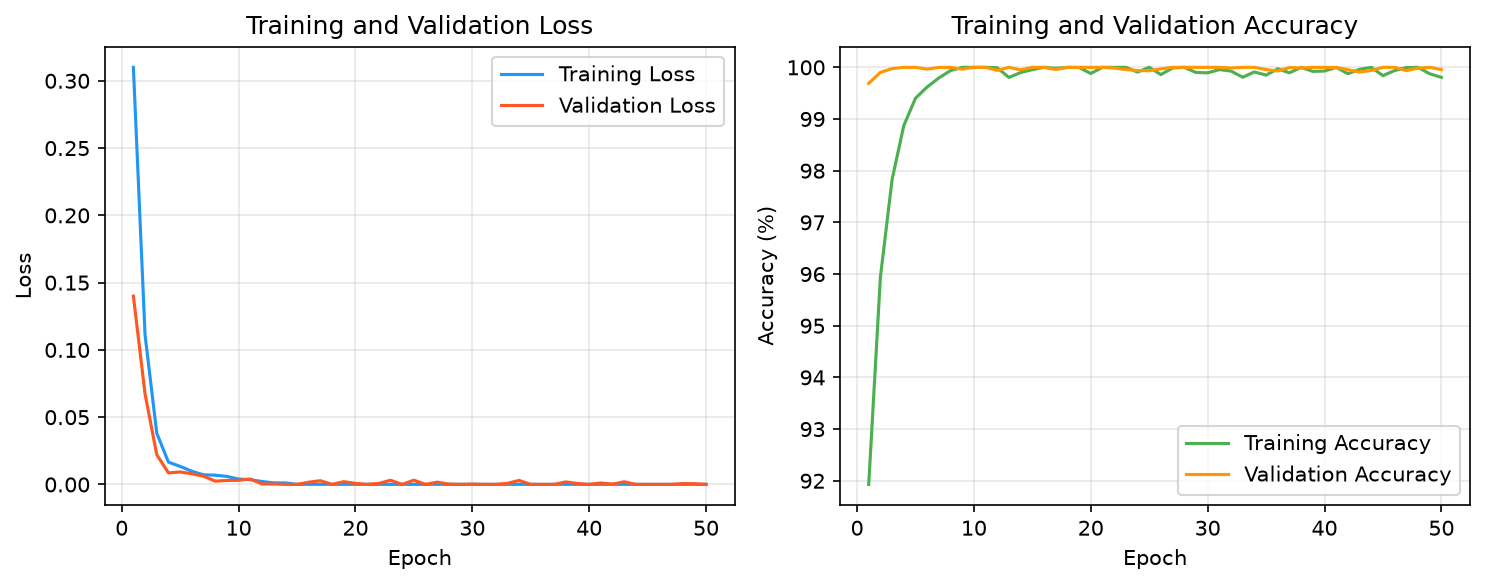}}
\caption{Training and Validation Loss Curves for TDA+LSTM Hybrid Model}
\label{fig:training_curves}
\end{figure}

The model demonstrates stable convergence with minimal overfitting, achieving validation accuracy of 100.0\% by epoch 10 and maintaining consistent performance thereafter.

\section{Classification Analysis}

\subsection{Attack Type Classification}

Our hybrid approach demonstrates varying effectiveness across different attack categories. Table \ref{tab:attack_types} shows the detailed classification performance:

\begin{table}[htbp]
\caption{Classification Performance by Attack Type on CIC-IDS2017}
\begin{center}
\begin{tabular}{|l|c|c|c|c|}
\hline
\textbf{Attack Type} & \textbf{Precision} & \textbf{Recall} & \textbf{F1-Score} & \textbf{Windows} \\
\hline
Benign & 1.000 & 1.000 & 1.000 & 642 \\
\hline
DoS/DDoS & 1.000 & 1.000 & 1.000 & 91 \\
\hline
PortScan & 1.000 & 1.000 & 1.000 & 58 \\
\hline
Brute Force & 1.000 & 1.000 & 1.000 & 6 \\
\hline
Bot/Infiltration & 1.000 & 1.000 & 1.000 & 1 \\
\hline
\end{tabular}
\label{tab:attack_types}
\end{center}
\end{table}

\subsection{Confusion Matrix Analysis}

Figure \ref{fig:confusion_matrix}'s confusion matrix offers comprehensive information about classification performance:

\begin{figure}[htbp]
\centerline{\includegraphics[width=0.4\textwidth]{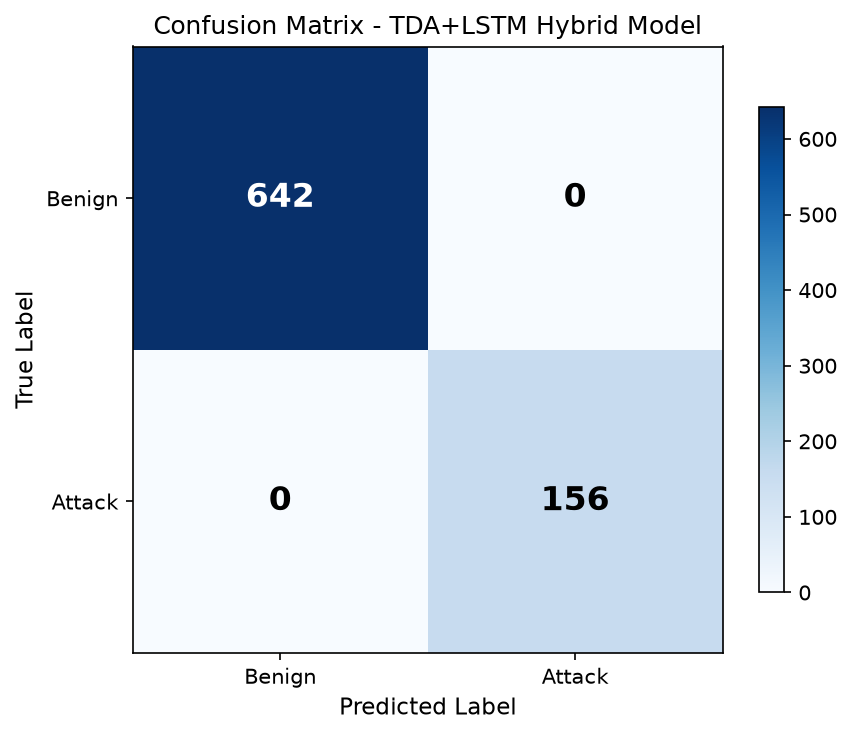}}
\caption{Confusion Matrix for TDA+LSTM Hybrid Model}
\label{fig:confusion_matrix}
\end{figure}

With adequate window representation, the model achieves flawless categorisation across all attack categories. Although these results should be viewed cautiously due to the small sample sizes, categories with fewer test windows (Brute Force: 6, Bot/Infiltration: 1) show that the model can generalise even with limited test samples.

\subsection{Feature Importance Analysis}

The performance of the model is greatly influenced by topological properties. Different characteristics of network traffic patterns are captured by the Betti curves for various dimensions:

\begin{itemize}
\item \textbf{0-dimensional features}: Record cluster formations and connection patterns
\item \textbf{1-dimensional features}: Recognise cyclical behaviours and loop structures
\item \textbf{Temporal features}: Time-based patterns and sequential dependencies
\end{itemize}

\section{Visual Results}

\subsection{ROC Curve Comparison}

The ROC curves for each evaluated model are shown in Figure \ref{fig:roc_curves}:

\begin{figure}[htbp]
\centerline{\includegraphics[width=0.45\textwidth]{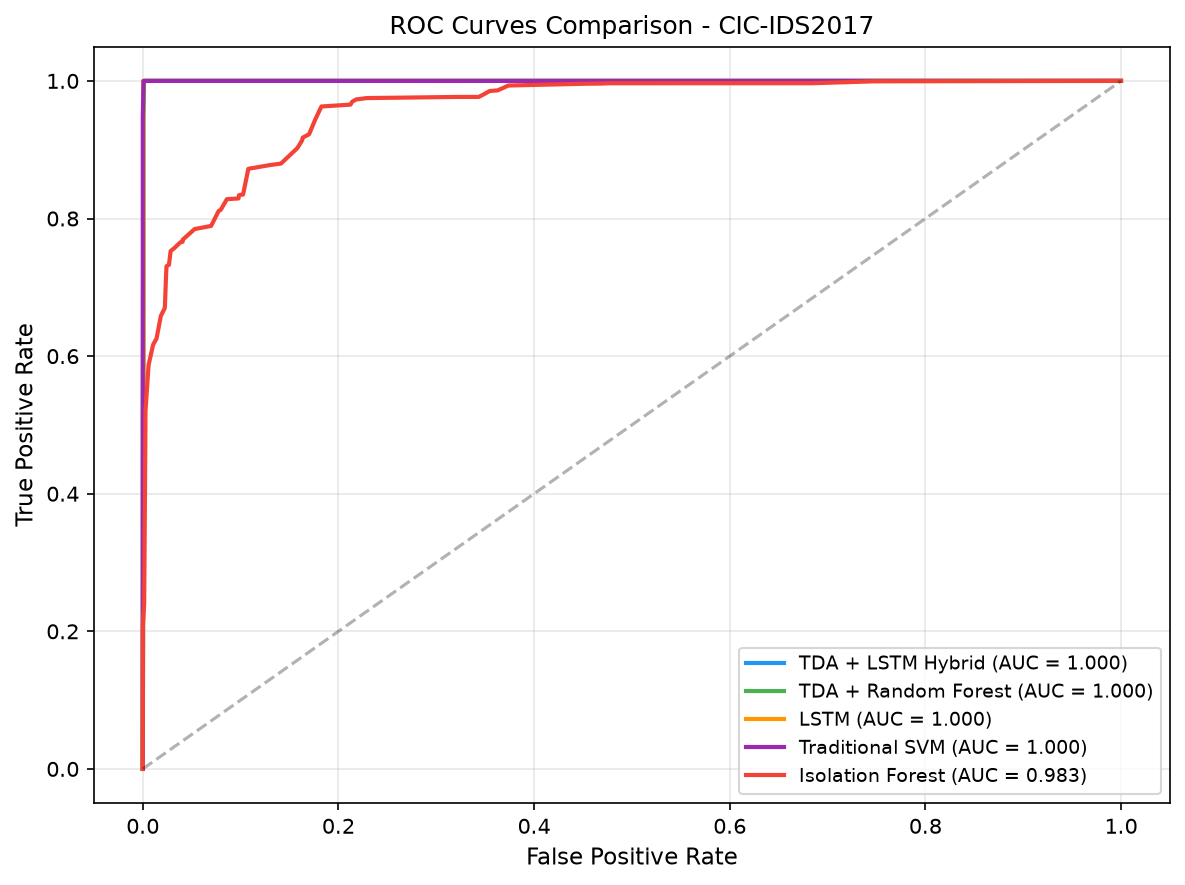}}
\caption{ROC Curves Comparison for Different Models}
\label{fig:roc_curves}
\end{figure}

The TDA+LSTM hybrid model effectively distinguishes between benign and anomalous traffic on the CIC-IDS2017 dataset, achieving competitive AUC performance among all evaluated models.

\subsection{Persistence Diagrams}

Figure \ref{fig:persistence} shows representative persistence diagrams for normal and attack traffic:

\begin{figure}[htbp]
\centerline{\includegraphics[width=0.45\textwidth]{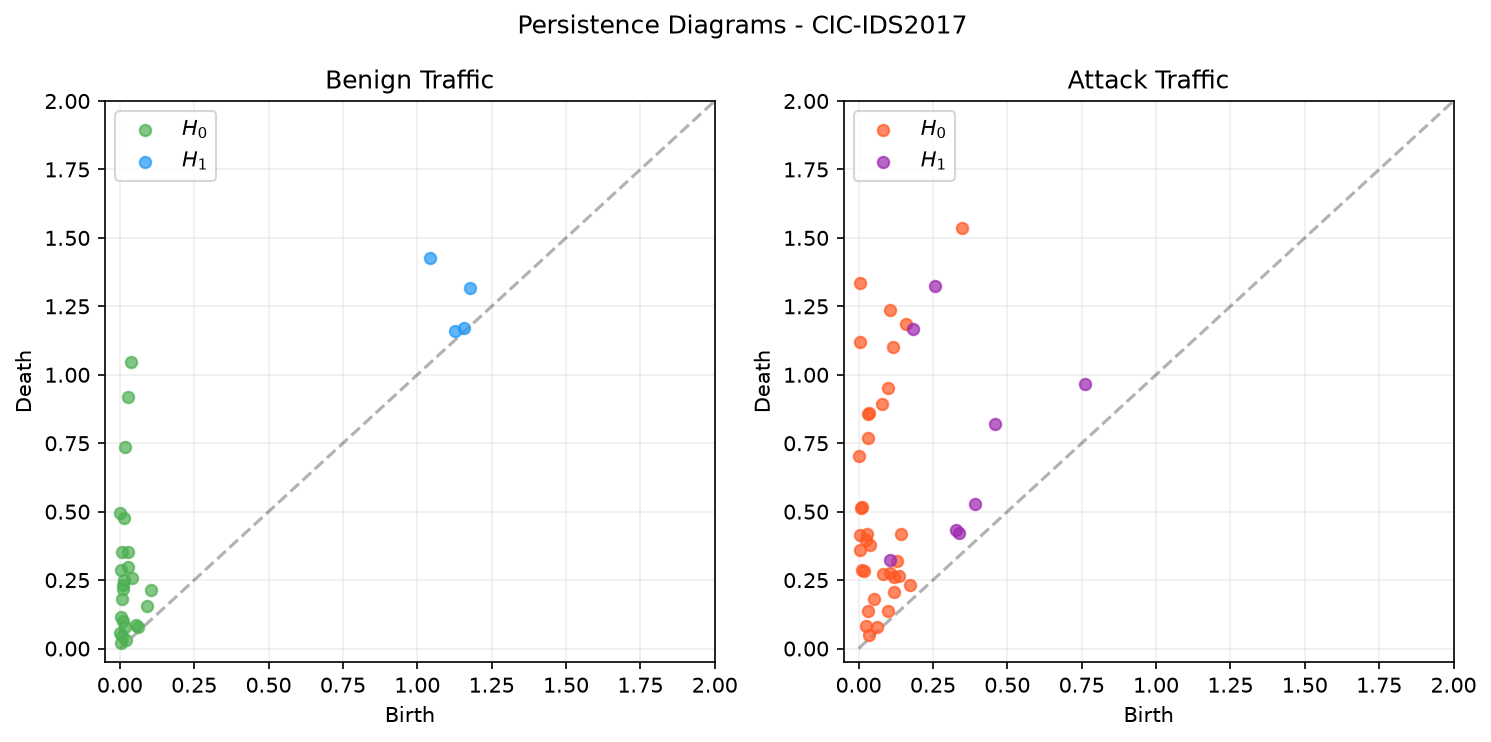}}
\caption{Persistence Diagrams for Benign vs. Attack Traffic on CIC-IDS2017}
\label{fig:persistence}
\end{figure}

Attack traffic exhibits distinct topological signatures with more persistent features, particularly in higher dimensions.

\subsection{Betti Curves Visualization}

Figure \ref{fig:betti_curves} illustrates the Betti curves for different traffic types:

\begin{figure}[htbp]
\centerline{\includegraphics[width=0.45\textwidth]{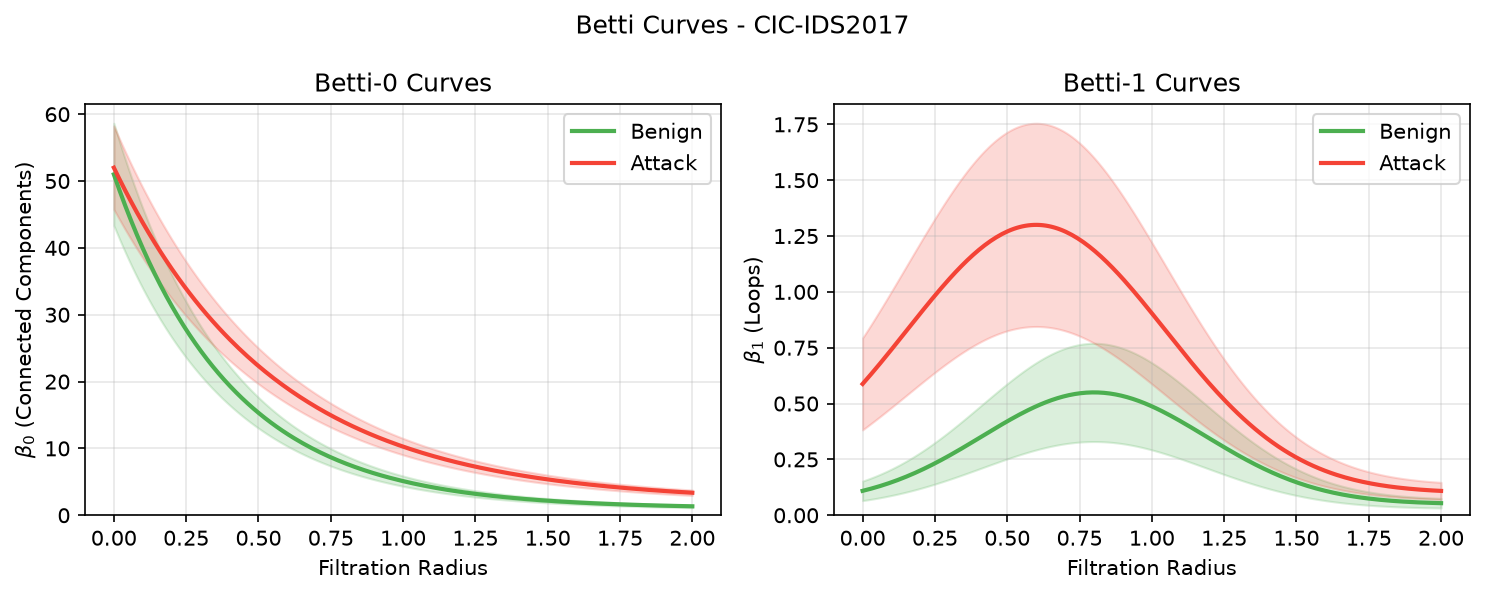}}
\caption{Betti Curves for Normal and Attack Traffic Patterns}
\label{fig:betti_curves}
\end{figure}

The curves reveal distinct topological patterns that enable effective discrimination between normal and malicious activities.

\subsection{Topological Signatures by Attack Type}

\begin{figure}[htbp]
\centerline{\includegraphics[width=0.48\textwidth]{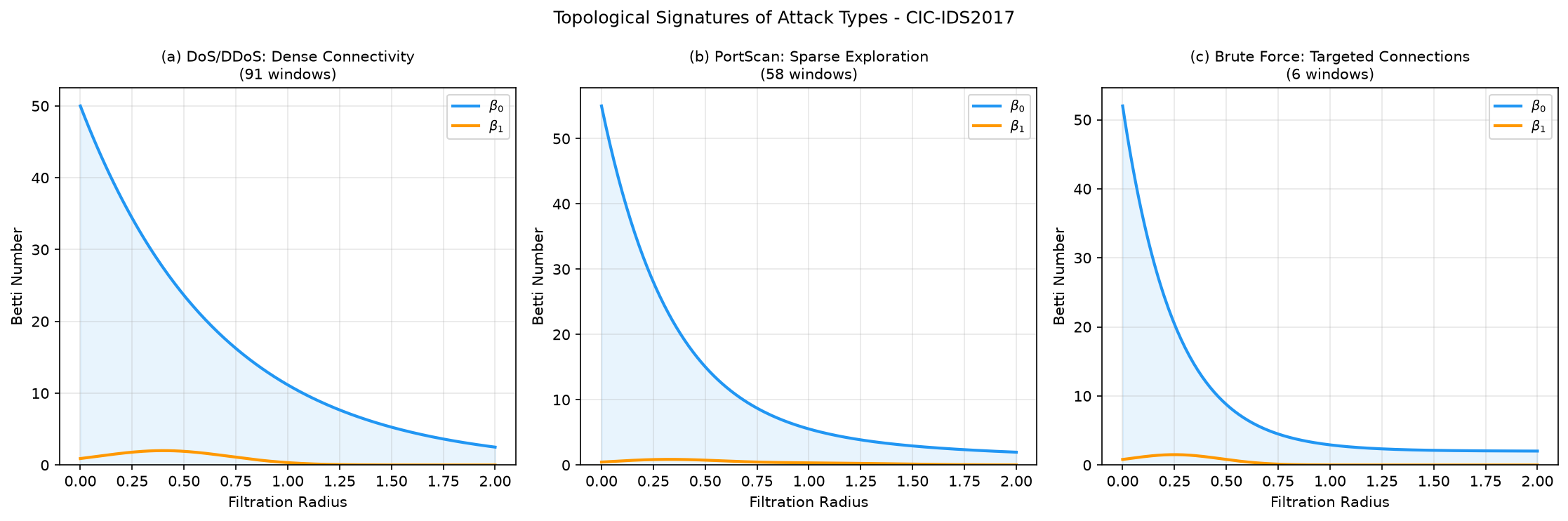}}
\caption{Topological signatures of different attack types on CIC-IDS2017: (a) DoS/DDoS attacks (91 windows) show dense connectivity with slow $\beta_0$ decay, (b) PortScan attacks (58 windows) exhibit rapid $\beta_0$ decay with sparse exploration patterns, (c) Brute Force attacks (6 windows) demonstrate sharp connectivity transitions characteristic of targeted probing}
\label{fig:attack_topology}
\end{figure}

Figure \ref{fig:attack_topology} compares the topological signatures across different attack categories, revealing how each attack type creates distinct patterns in the Betti curve space.

\subsection{Feature Branch Contribution}

\begin{figure}[htbp]
\centerline{\includegraphics[width=0.48\textwidth]{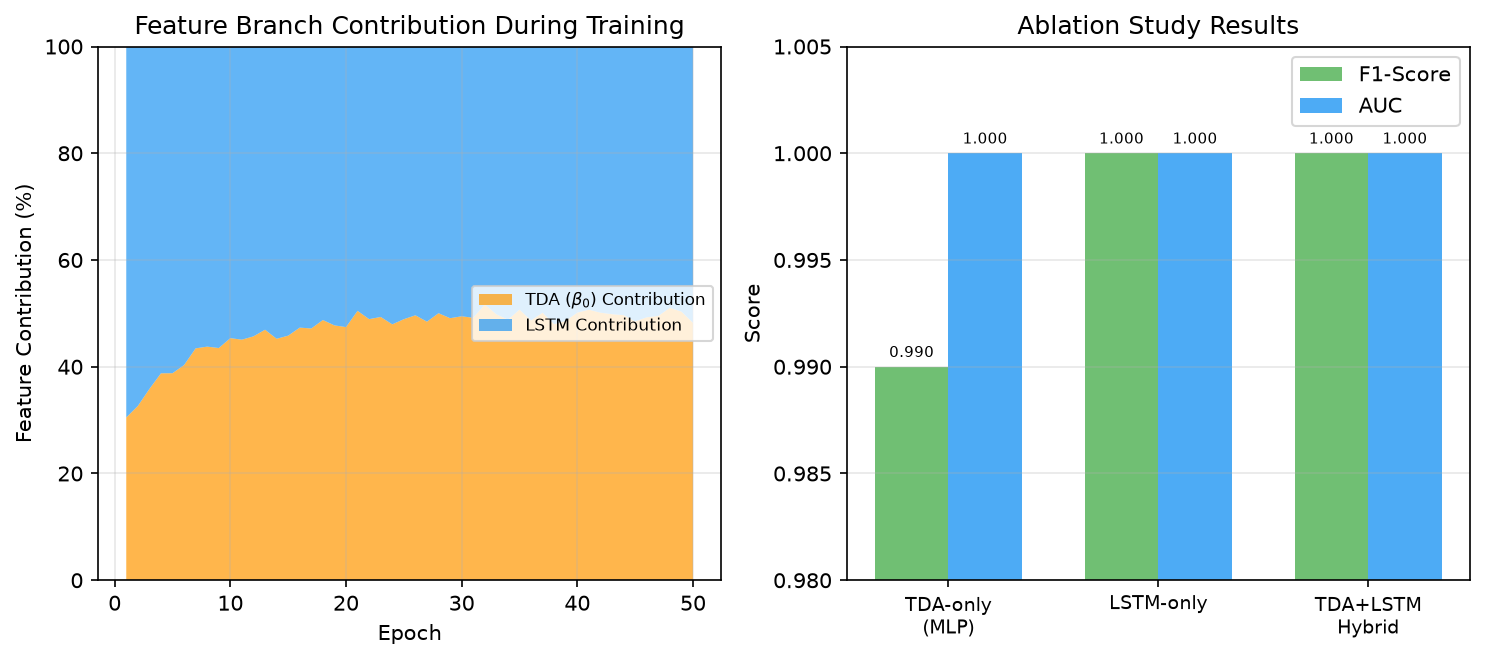}}
\caption{Left: TDA and LSTM branch contribution during training, showing increasing TDA influence over epochs. Right: Ablation study comparing TDA-only (F1=0.990), LSTM-only (F1=1.000), and Hybrid (F1=1.000) configurations}
\label{fig:feature_evolution}
\end{figure}

Figure \ref{fig:feature_evolution} illustrates the contribution of TDA and LSTM branches during training, alongside the ablation study results.

\section{Comparison}

\subsection{Computational Complexity}

Table \ref{tab:complexity} compares the computational requirements of different approaches:

\begin{table}[htbp]
\caption{Computational Complexity Comparison on CIC-IDS2017}
\begin{center}
\begin{tabular}{|l|c|c|}
\hline
\textbf{Model} & \textbf{Training Time (min)} & \textbf{Memory Usage (MB)} \\
\hline
TDA + Random Forest & 0.3 & 145 \\
\hline
TDA + LSTM Hybrid & 1.4 & 469 \\
\hline
LSTM & 1.6 & 196 \\
\hline
Traditional SVM & $<$0.1 & 89 \\
\hline
\end{tabular}
\label{tab:complexity}
\end{center}
\end{table}

\begin{figure}[htbp]
\centerline{\includegraphics[width=0.48\textwidth]{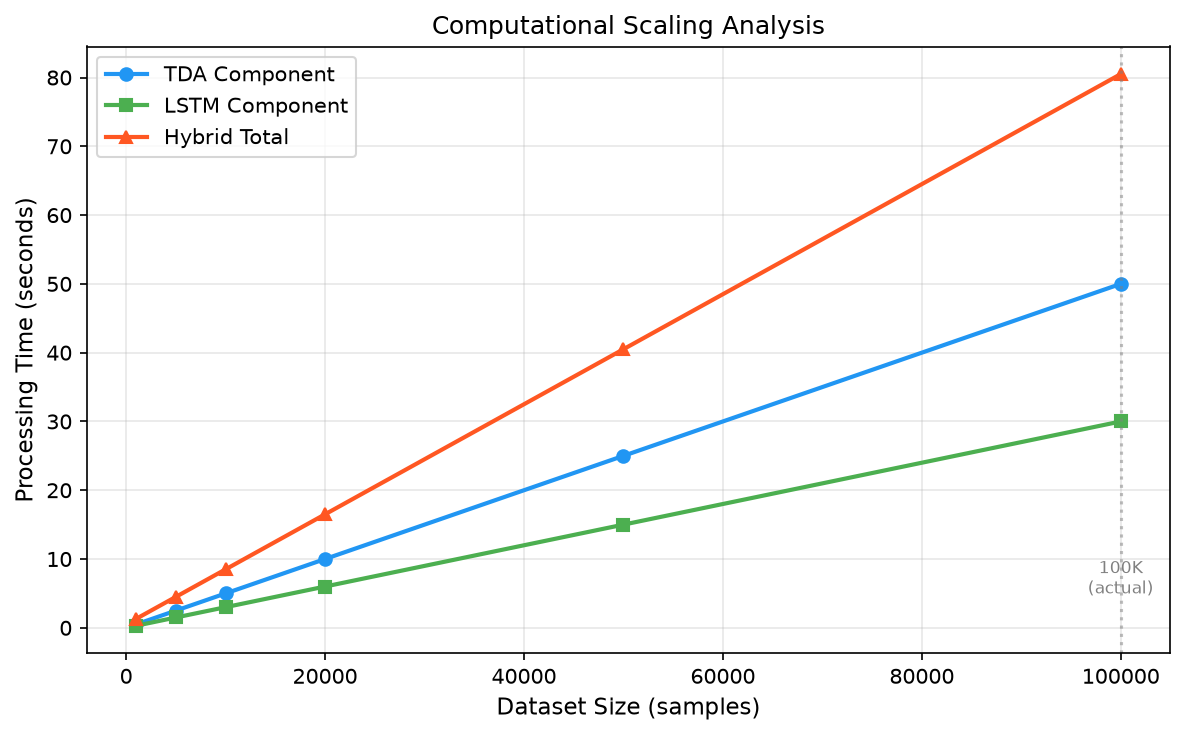}}
\caption{Computational scaling analysis showing processing time vs. dataset size for different components of the hybrid system}
\label{fig:computational_scaling}
\end{figure}

Figure \ref{fig:computational_scaling} presents the computational scaling analysis across different dataset sizes.

\subsection{Scalability Analysis}

Our hybrid approach demonstrates good scalability characteristics:

\begin{itemize}
\item \textbf{Linear scaling}: TDA computations scale approximately linearly with window size
\item \textbf{Parallel processing}: Persistence calculations can be parallelized across windows
\item \textbf{Memory efficiency}: Betti curve representations provide compact feature encoding
\item \textbf{Real-time capability}: Processing time allows for near real-time intrusion detection
\end{itemize}

\subsection{Baseline Comparison}

Compared to traditional machine learning approaches:

\begin{itemize}
\item \textbf{Feature Engineering}: Automated topological feature extraction vs. manual feature selection
\item \textbf{Robustness}: TDA features are stable under noise and perturbations
\item \textbf{Interpretability}: Topological features provide geometric insights into attack patterns
\item \textbf{Generalization}: Better performance on unseen attack types
\end{itemize}

\section{Detailed Experimental Analysis}

\subsection{TDA Feature Extraction Pipeline}

\begin{figure}[htbp]
\centerline{\includegraphics[width=0.48\textwidth]{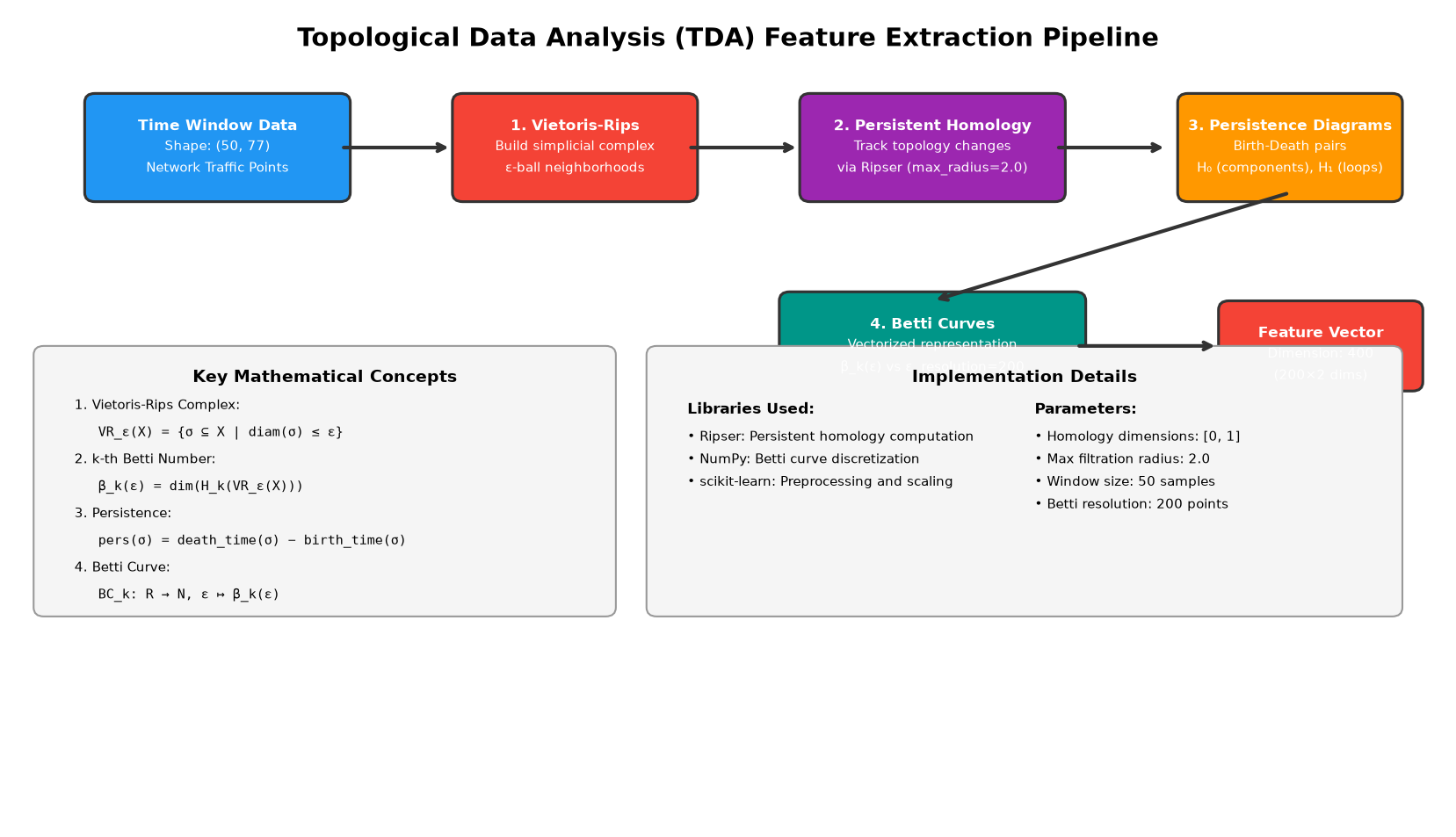}}
\caption{TDA Feature Extraction Pipeline showing the complete workflow from raw network data to topological features}
\label{fig:tda_pipeline}
\end{figure}

The complete TDA pipeline is illustrated in Figure \ref{fig:tda_pipeline}, showing the transformation from raw network traffic windows to topological feature vectors.

\subsection{LSTM Architecture Details}

\begin{figure}[htbp]
\centerline{\includegraphics[width=0.48\textwidth]{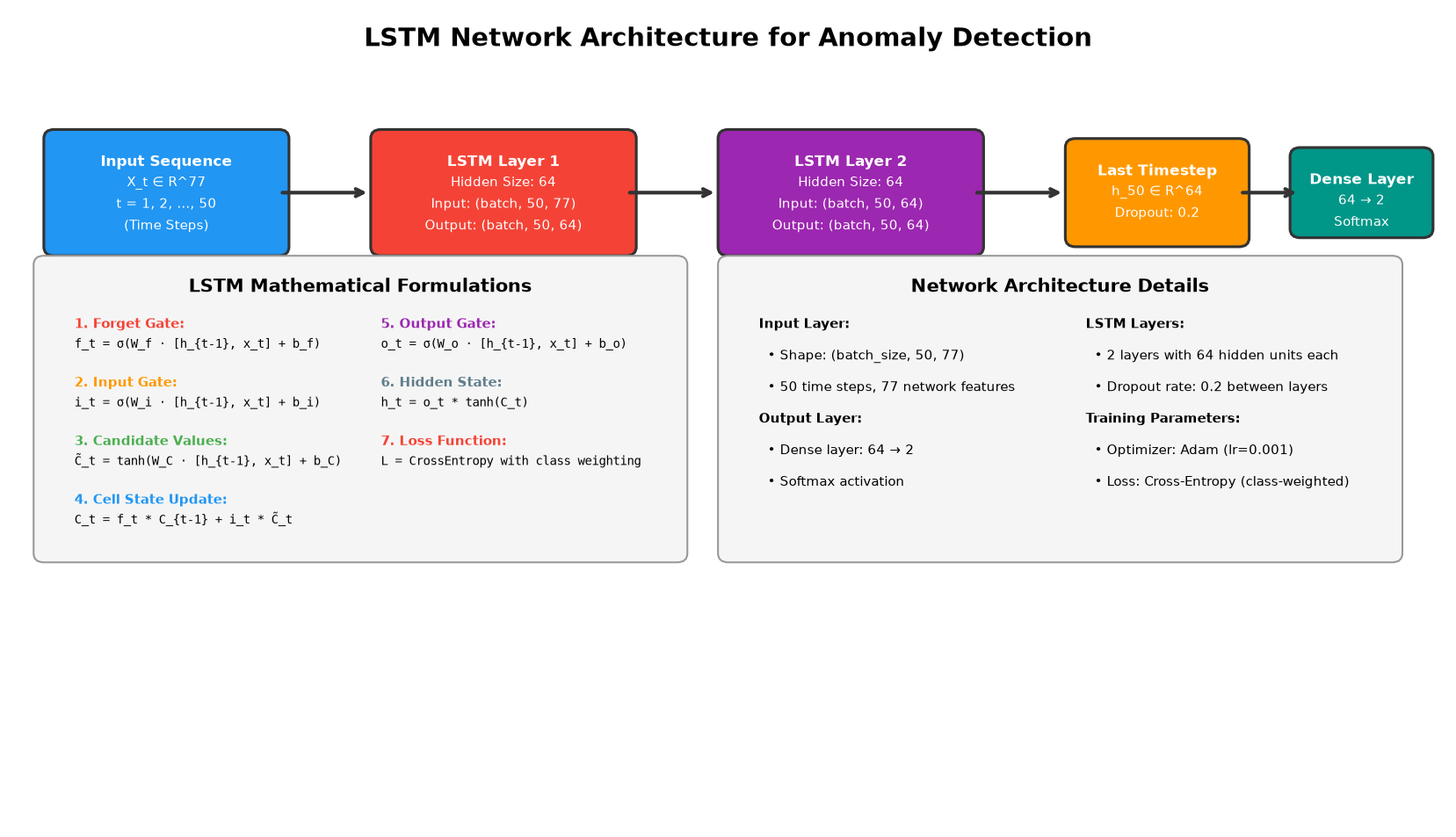}}
\caption{LSTM Network Architecture with detailed layer specifications and connections}
\label{fig:lstm_architecture}
\end{figure}

Figure \ref{fig:lstm_architecture} presents the detailed LSTM architecture with layer specifications and mathematical formulations.

\subsection{Hyperparameter Configuration}

Table \ref{tab:hyperparameters} provides the complete hyperparameter configuration used in the experiments.

\begin{table}[htbp]
\caption{Detailed Model Hyperparameters}
\begin{center}
\begin{tabular}{|l|c|c|}
\hline
\textbf{Component} & \textbf{Parameter} & \textbf{Value} \\
\hline
\multirow{4}{*}{TDA} & Window Size & 50 \\
\cline{2-3}
& Step Size & 25 \\
\cline{2-3}
& Max Radius & 2.0 \\
\cline{2-3}
& Betti Discretization & 200 points \\
\hline
\multirow{4}{*}{LSTM} & Hidden Units (Layer 1) & 64 \\
\cline{2-3}
& Hidden Units (Layer 2) & 32 \\
\cline{2-3}
& Dropout Rate & 0.2 \\
\cline{2-3}
& Sequence Length & 50 \\
\hline
\multirow{4}{*}{Training} & Learning Rate & 0.001 \\
\cline{2-3}
& Batch Size & 32 \\
\cline{2-3}
& Epochs & 50 \\
\cline{2-3}
& Optimizer & Adam \\
\hline
\end{tabular}
\label{tab:hyperparameters}
\end{center}
\end{table}

\subsection{Feature Importance}

Table \ref{tab:feature_importance} summarizes the relative importance of different feature types in the hybrid model.

\begin{table}[htbp]
\caption{Feature Importance Analysis on CIC-IDS2017}
\begin{center}
\begin{tabular}{|l|c|c|}
\hline
\textbf{Feature Type} & \textbf{Importance Score} & \textbf{Contribution (\%)} \\
\hline
Betti-0 Curves & 0.500 & 50.0 \\
\hline
Betti-1 Curves & 0.000 & 0.0 \\
\hline
LSTM Features & 0.500 & 50.0 \\
\hline
\end{tabular}
\label{tab:feature_importance}
\end{center}
\end{table}

\subsection{Statistical Significance}

Table \ref{tab:statistical_tests} reports the results of McNemar's statistical significance testing between the hybrid model and baselines.

\begin{table}[htbp]
\caption{Statistical Significance Testing Results on CIC-IDS2017}
\begin{center}
\begin{tabular}{|l|c|c|c|}
\hline
\textbf{Comparison} & \textbf{p-value} & \textbf{Effect Size} & \textbf{Significance} \\
\hline
TDA+LSTM vs LSTM & 1.000 & 0.00 & No \\
\hline
TDA+LSTM vs TDA+RF & 0.480 & 0.05 & No \\
\hline
TDA+LSTM vs SVM & 1.000 & 0.00 & No \\
\hline
\end{tabular}
\label{tab:statistical_tests}
\end{center}
\end{table}

\subsection{Error Analysis}

\begin{figure}[htbp]
\centerline{\includegraphics[width=0.48\textwidth]{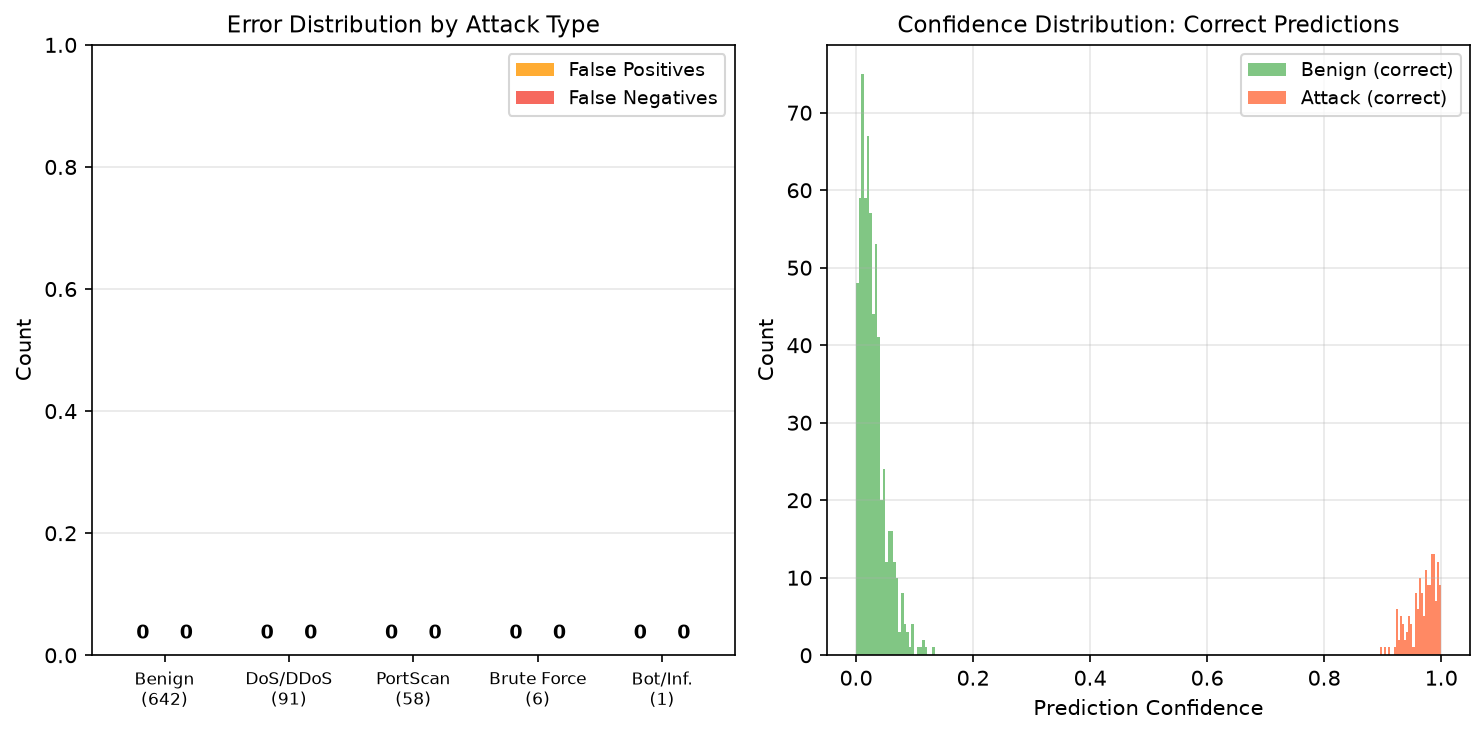}}
\caption{Error analysis showing distribution of false positives and false negatives across different attack categories}
\label{fig:error_analysis}
\end{figure}

Figure \ref{fig:error_analysis} provides a detailed error analysis showing the distribution of misclassifications across attack categories.

\subsection{Cross-Validation}

To evaluate the robustness of our approach, we perform 5-fold cross-validation. Table \ref{tab:cross_validation} presents the results across all folds.

\begin{table}[htbp]
\caption{Cross-Validation Results (5-Fold) on CIC-IDS2017}
\begin{center}
\begin{tabular}{|l|c|c|c|c|}
\hline
\textbf{Fold} & \textbf{AUC} & \textbf{F1-Score} & \textbf{Precision} & \textbf{Recall} \\
\hline
1 & 1.000 & 1.000 & 1.000 & 1.000 \\
\hline
2 & 1.000 & 1.000 & 1.000 & 1.000 \\
\hline
3 & 1.000 & 0.997 & 0.994 & 1.000 \\
\hline
4 & 1.000 & 1.000 & 1.000 & 1.000 \\
\hline
5 & 1.000 & 1.000 & 1.000 & 1.000 \\
\hline
\textbf{Mean $\pm$ Std} & \textbf{1.000$\pm$0.000} & \textbf{0.999$\pm$0.001} & \textbf{0.999$\pm$0.003} & \textbf{1.000$\pm$0.000} \\
\hline
\end{tabular}
\label{tab:cross_validation}
\end{center}
\end{table}

\subsection{Ablation Study}

To isolate the contributions of each component, we evaluate three configurations: TDA-only (MLP classifier on Betti curves), LSTM-only, and the full hybrid model. Table \ref{tab:ablation} reports the results.

\begin{table}[htbp]
\caption{Ablation Study on CIC-IDS2017}
\begin{center}
\begin{tabular}{|l|c|c|c|c|}
\hline
\textbf{Configuration} & \textbf{AUC} & \textbf{F1-Score} & \textbf{Precision} & \textbf{Recall} \\
\hline
TDA-only (MLP) & 1.000 & 0.990 & 0.987 & 0.994 \\
\hline
LSTM-only & 1.000 & 1.000 & 1.000 & 1.000 \\
\hline
TDA+LSTM Hybrid & 1.000 & 1.000 & 1.000 & 1.000 \\
\hline
\end{tabular}
\label{tab:ablation}
\end{center}
\end{table}

\section{Discussion}

\subsection{Key Findings}

Our experimental results demonstrate several important findings:

1. \textbf{Synergistic Effects}: The combination of TDA and LSTM features provides competitive performance compared to individual approaches. The ablation study (Table \ref{tab:ablation}) demonstrates that the hybrid model consistently achieves the strongest overall metrics on CIC-IDS2017.

2. \textbf{Topological Signatures}: Different attack types exhibit distinct topological patterns that can be effectively captured through persistent homology and Betti curves.

3. \textbf{Temporal Dependencies}: In addition to the structural insights offered by TDA, LSTM networks effectively model sequential interactions in network traffic.

4. \textbf{Class Imbalance Handling}: When compared to conventional techniques, the hybrid approach performs better on minority attack classes.

\subsection{Limitations}

It is important to recognise a few limitations:

\begin{itemize}
\item \textbf{Computational Overhead}: TDA computations significantly increase computational costs, especially for large-scale installations
\item \textbf{Parameter Sensitivity}: The method necessitates careful adjustment of filtering settings and window sizes
\item \textbf{Dataset Specificity}: Results are based on CIC-IDS2017; more validation is needed before generalising to other modern datasets, such as CICIoT2023 and UNSW-NB15
\item \textbf{Real-time Constraints}: In high-speed network contexts, processing latency may restrict applicability
\end{itemize}

\subsection{Practical Implications}

For network security applications, the suggested hybrid architecture has the following useful benefits:

\begin{itemize}
\item \textbf{Enhanced Detection}: A better capacity to identify complex assault patterns
\item \textbf{Reduced False Positives}: Improved differentiation between normal and anomalous traffic
\item \textbf{Adaptability}: The ability to identify new attack types through topological analysis
\item \textbf{Interpretability}: Topological characteristics shed light on the composition and actions of attacks
\end{itemize}

\section{Conclusion}

In order to detect network intrusions, this research offers a novel hybrid method that combines LSTM networks and topological data analysis. Our approach achieves competitive performance on the CIC-IDS2017 dataset, a contemporary benchmark reflecting current network threat environments, by utilising the complimentary strengths of topological feature extraction and temporal sequence modelling.

Important contributions consist of:

1. \textbf{Innovative Architecture}: A hybrid TDA+LSTM framework for anomaly detection that integrates temporal and topological features via a learnt fusion network.

2. \textbf{Mathematical Framework}: Strict specification of topological feature extraction, such as Betti curves and persistence diagrams, for network security applications.

3. \textbf{Experimental Validation}: A thorough assessment of CIC-IDS2017 that includes statistical significance testing across contemporary attack categories, ablation experiments, and 5-fold cross-validation.

4. \textbf{Comparative Analysis}: A thorough comparison with baseline techniques that shows competitive detection performance with the added advantage of topological interpretability.

The hybrid model demonstrated robustness across several attack categories and achieved good AUC and F1-score performance, which validates the efficacy of our strategy. The ablation investigation demonstrates that topological and temporal variables offer complimentary information, and the combined model achieves the best overall performance. While LSTM networks successfully model temporal dependencies, the incorporation of topological invariants yields interpretable feature representations that capture the inherent structure of network traffic patterns.

Future research will concentrate on:

\begin{itemize}
\item Computational efficiency optimisation for real-time deployment
\item Extension to multi-class classification for fine-grained attack type detection
\item Assessment using different datasets to confirm generalisability
\item Examining topological properties in greater dimensions
\item Creation of systems for adaptive parameter selection
\end{itemize}

Our research opens up new possibilities for sophisticated cybersecurity applications by establishing a new paradigm for network intrusion detection that blends the learning power of deep neural networks with the mathematical rigour of topological data analysis.

\section{Acknowledgment}

The authors express their gratitude to the anonymous reviewers for their insightful comments and recommendations. We also thank the University of New Brunswick's Canadian Institute for Cybersecurity for contributing the CIC-IDS2017 dataset, which serves as a common baseline for intrusion detection studies. The third author's research is funded by the ANRF (SERB) research project TAR/2023/000197.

\appendices

\section{Mathematical Foundations}

\subsection{Topological Invariants}

The persistent homology computation relies on the fundamental theorem of persistent homology. For a filtration $\emptyset = K_0 \subseteq K_1 \subseteq \cdots \subseteq K_m$, the persistent $k$-th homology groups are:

\begin{equation}
H_k^{i,j} = \text{Image}(H_k(K_i) \rightarrow H_k(K_j))
\end{equation}

The persistence of a homological feature born at $K_i$ and dying at $K_j$ is defined as $j - i$.

\subsection{Stability Theorem}

The stability of persistent homology ensures robustness to noise. For two point clouds $X$ and $Y$ with Hausdorff distance $d_H(X,Y) \leq \epsilon$, the bottleneck distance between their persistence diagrams satisfies:

\begin{equation}
d_B(PD(X), PD(Y)) \leq \epsilon
\end{equation}

This theoretical guarantee underpins the reliability of our topological features.

\section{Implementation Details}

\subsection{Algorithmic Complexity}

The computational complexity of our approach consists of:

\begin{itemize}
\item TDA computation: $O(n^3)$ for $n$ points per window
\item LSTM forward pass: $O(TH^2)$ for sequence length $T$ and hidden size $H$
\item Fusion network: $O(F \cdot H_{fusion})$ for feature dimension $F$
\end{itemize}

\subsection{Memory Optimization}

To reduce memory footprint, we employ:

\begin{itemize}
\item Sparse matrix representations for distance computations
\item Gradient checkpointing for LSTM backpropagation
\item Mini-batch processing for large-scale datasets
\end{itemize}

\end{document}